\documentclass[nologo,url,11pt,a4paper]{ETHpaper}

\usepackage{url}
\usepackage{stfloats}
\usepackage{color}
\usepackage{graphicx}
\usepackage{subcaption}
\usepackage{dcolumn}
\usepackage{booktabs}
\usepackage{multirow}
\usepackage{amssymb}
\usepackage{amsmath}
\usepackage{amsfonts}

\newcommand{\mean}[1]{\left\langle #1 \right\rangle}

\newcommand{\yt}{\texttt{YouTube}}
\newcommand{\red}{\texttt{Reddit}}
\newcommand{\ud}{\texttt{Urban Dictionary}}
\newcommand{\img}{\texttt{Imgur}}
\newcommand{\tw}{\texttt{Twitter}}
\newcommand{\fb}{\texttt{Facebook}}
\newcommand{\posit}{\textbf{P}}
\newcommand{\negit}{\textbf{N}}
\newcommand{\arousal}{\textbf{A}}
\newcommand{\valence}{\textbf{V}}
\newcommand{\dominance}{\textbf{D}}
\newcommand{\like}{\texttt{like}}
\newcommand{\dislike}{\texttt{dislike}}
\newcommand{\rlang}{\textbf{\textsf{R}}}

\begin{document}

\title{When the Filter Bubble Bursts:\\
Collective Evaluation Dynamics in Online Communities}

\titlealternative{When the Filter Bubble Bursts: Collective Evaluation Dynamics 
in Online Communities\\
\textbf{Submitted to the 8th International ACM Web Science Conference 2016}}

\author{Adiya Abisheva, David Garcia, Frank Schweitzer}

\maketitle

\begin{abstract} 

We analyze online collective evaluation processes through positive and
negative votes in various social media.   We find two modes of collective
evaluations that stem from the existence of  filter bubbles.  Above a
threshold of collective attention, negativity grows faster with positivity, as
a sign of the burst of a filter bubble when information reaches beyond the
local social context of a user. We analyze how collectively evaluated content
can reach  large social contexts and create polarization, showing that
emotions expressed through text play a key role in collective evaluation
processes.

\end{abstract}

\small {
\textbf{Categories and Subject Descriptors:} Human-centered computing, 
Collaborative and social computing, Empirical studies in collaborative and 
social computing

\textbf{Keywords:} Social filtering, emotions, collective dynamics}

\section{Introduction}

\paragraph{When the filter bubble bursts}

Rebecca Black, an amateur teenage singer, posted a music video\footnotemark[1] 
on \yt\ on February 10, 2011. The song originally circulated mostly among the 
Facebook friends of its 13-year old singer and was loved and positively 
commented. Rebecca Black's song received the "all  the usual friends things" 
\cite{Larsen2011} and was enough to please her, but it suddenly went viral 
\textit{in the wrong direction}. From initial 4,000 views on \yt\, her song
skyrocketed to 13  Million views. This sudden popularity brought mostly
negative attention, up to the point of becoming officially the most disliked 
\yt\ video\footnotemark[2], and by  June 15, 2011 the song received 3.2 Million 
dislikes in \yt\ against less than half a million likes. From \textit{local} 
fame her song soared to the heights of \textit{global} shame.



\begin{figure}[h]
  \centering
  \includegraphics[width=0.68\textwidth]{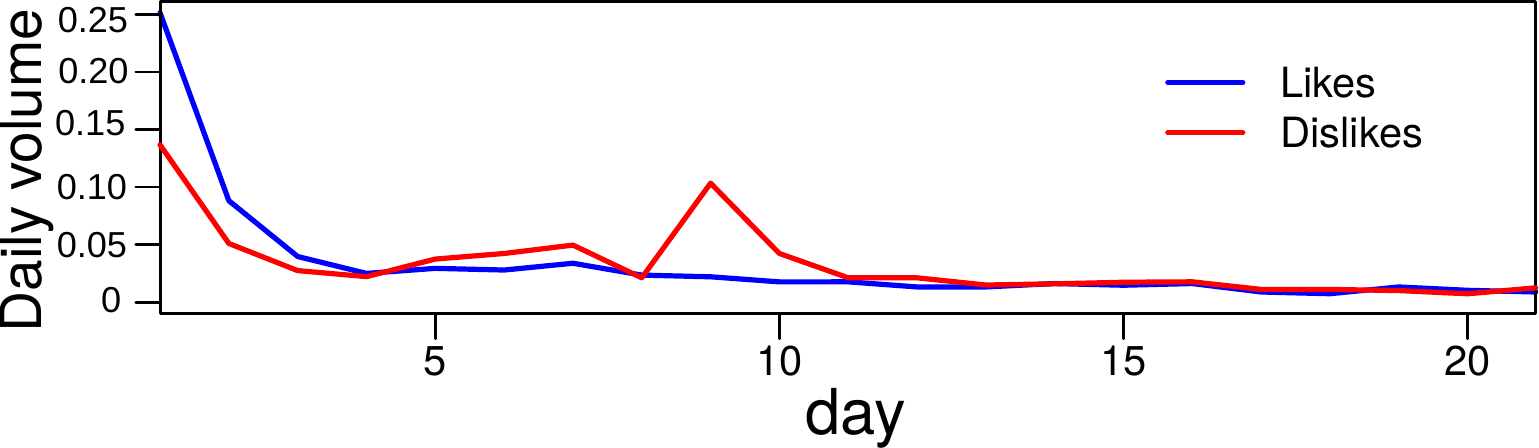}
  \caption{\textbf{Example of evaluation dynamics in Youtube.} 
  Normalized daily volume of likes and dislikes for a video in our \yt\ dataset.
Likes appear soon after the video is uploaded, while dislikes tend to appear 
later.}
  \label{fig:example}
  \vspace{-7pt}
\end{figure}

The anecdotal example of Rebecca Black's song is paradigmatic of some aspects
of the collective dynamics of evaluations in online media. A video can become 
relatively popular within a small community and receive initial positive 
evaluations, but when larger audiences are reached, negativity rises faster than 
in early moments. Figure \ref{fig:example} shows this phenomenon through an 
example of the relative daily volume of likes and dislikes of a \yt\ video. 
Initially, the video is positively evaluated, but the volume of likes decreases 
quickly. While initial dislikes also decrease, they start rising after the 
fourth day, reaching a peak at the ninth day.

\footnotetext[1]{The original video was deleted and reuploaded again at:
\url{https://www.youtube.com/watch?v=kfVsfOSbJY0}}
\footnotetext[2]{\small \url{http://knowyourmeme.com/memes/rebecca-black-friday}}

The early viewers of a \yt\ video are prone to like it, either due to a social
connection with the uploader, or given the similarity of the video with their
past liked content. This is a consequence of the purpose of social filtering
mechanisms and recommender systems, which is to personalize content selection
such that users find content that they consider relevant and of good quality.
In contrast, the video can also spread through other media towards more
general users, and eventually reach a global audience with users more critical
or negative towards the video. Beyond \yt\ videos, this phenomenon can be seen
as another aspect of the filter bubble \cite{Pariser2011}: The reinforcement
of opinions caused by filtering mechanisms creates an initial pocket of
positivity, but \emph{when the filter bubble bursts}, collective negativity
can backlash.

Our study sets out to understand collective evaluation processes in various
social media through likes and dislikes, as manifestations of opinions towards
the evaluated content.  We test the duality of collective evaluations in the
local versus global behavior illustrated above, looking for the existence of a
threshold of positivity after which negative evaluations rise faster and
polarization emerges.

\paragraph{Emotions in polarization}

Technological filters are not the only factor that shapes collective
evaluations; emotions influence cognitive information processing, shaping
opinions and attitudes towards online content. A number of studies in social
psychology show how emotions influence  individual evaluations, judgements,
and opinions \cite{Kuhne2012,Gonzalez-Bailon2010,Gorn2011}, based on the
theory of \emph{core affect} \cite{Russell1999}. Within this theoretical
framework, emotions are composed of two dimensions: i) \textit{valence}, which
characterizes the feeling of pleasure or displeasure, and  ii) \textit{arousal}, 
which encompasses a feeling of activation or deactivation, and quantifies  
mobilization and energy \cite{Russell1999}. Additional dimensions can improve 
the representation of emotional experience, such as potency or surprise 
\cite{Fontaine2007}, but their consistent inclusion in psychological research 
about opinions is  still to be explored.

Research in psychology on the role of emotions in evaluations show that arousal 
can lead to extreme reactions and polarized responses \cite{Reisenzein1983}.    
The theory of misattribution explains this effect \cite{Zillmann1971, 
Reisenzein1983} as a transfer of residual emotions between events that 
intensifies the reaction to the second event.  For instance, men in a state of 
high emotional arousal (for example from physical exercises) give more extreme 
ratings of attractiveness to women in comparison to the situation in which 
raters are in a calm emotional state. Similarly, valence can be misattributed 
and bias evaluations \cite{Schwarz1988}, in particular when individuals 
inspecting their  current feelings, which might be caused by an incidental 
source rather than the evaluated content.

Further theoretical explanations for the role of emotions in evaluations pose
the reduction of cognitive complexity induced by emotional states, which bias
the formulation of evaluations towards fast rather than informed responses.
The theory of affect priming explains this through the attribution of an
individual's mood to  similarly valenced signals in memory, which helps
reducing the effort of evaluation tasks \cite{Isen1978}. Empirical evidence
shows that the subjective experience of arousal  motivates evaluation on the
extremes \cite{Paulhus1994}. For example,  the ratings of famous figures by
students are found to be more polarized right before taking an exam, in
comparison to weeks before or after. This kind of reactions are especially
salient when arousal is experienced along with negative valence (such as the
stress before an exam), and thus  we can expect the expression of negative and
aroused emotions to motivate  more polarized collective evaluations in social
media. The digital traces of collective evaluations allow us to analyze
further the role of  emotions in online evaluation processes.


\paragraph{Contributions of this article}

In this work we analyze collective evaluations across different social media
to reveal statistical regularities related to information filters and
emotions. First, we test if the distributions of likes and dislikes of
evaluated content shows signs of the existence of multiplicative growth
processes of social interaction. Second, we test if the relationship between
likes and dislikes is non-linear with a division in to two different modes,
corresponding to  local and global collective evaluations. Third, we test if
the emotions expressed in the evaluated content lead to global and polarized
collective responses.  Our work provides insights into the properties of
collective evaluations and tests established psychology theories on the role
of emotions in opinion formation.

\section{Background}

\paragraph{Collective dynamics}

In the last years, lots of research focused on the topic of online
communities,  i.e. large groups of individuals that interact through an online
medium. Collective  phenomena such as dynamics of trends
\cite{Wu2007,Wang2012}, or viral  marketing \cite{Leskovec2007} can be
assessed with data from online  communities.   Examples of studies on online
user behaviour are understanding dynamics  of replying activity and website
engagement of users  \cite{Rowe2014}, buyer activity in online shopping
websites \cite{Lee2015}  and communication dynamics in forums \cite{Kan2011}.
Another example is  research on social influence, which was shown to exist in
\yt\ \cite{Crane2008}, in \fb\ \cite{Onnela2010}, and in \tw\
\cite{Varol2014}. Furthermore, social influence on popularity of 
\texttt{Facebook} applications has been shown to arise from a mixture of local 
and global signals \cite{Onnela2010}. While the former notion indicates how 
friends and local community influence an individual's behaviour, the latter 
suggests the effect of the aggregate popularity of products or behaviours on an 
individual. Additionally, previous results for popularity distributions show 
that the amount of votes for \texttt{Digg} stories \cite{Mieghem2011b} and 
tweets in trending topics \cite{Asur2011} follow log-normal distributions that 
are explained by social coupling.

\paragraph{Collective evaluations}
Online voting dynamics and dynamics of human appraisal were studied in a number 
of previous research. Studies on collective evaluations mostly interrelates and 
finds explanations in research on collective popularity of the online content, 
with the assumption that more likes lead to item's popularity. Despite the 
differences in the ways of measuring popularity, as a number of views in \yt\ 
\cite{Szabo2010}, or as a number of likes and dislikes in \red\ 
\cite{Mieghem2011a}, or as a number of votes in \texttt{Digg} 
\cite{Mieghem2011b} or as a time span of trending topics in \tw\ 
\cite{Asur2011}, these measures showed the existence of statistical 
regularities of content popularity, and fit to the log-normal distribution.

Studies on collective dynamics of negative evaluations are scarcer, but some
recent works illustrate that social influence effects are present  in movie
ratings from \texttt{imdb.com} \cite{Lorenz2009}, and that controversiality
expressed through movie ratings evolves with time \cite{Amendola2015}.
Additionally, herding effects have been observed in random manipulations of
votes  in \red\ \cite{Weninger2015}, which shows that the way users vote
depends on  the votes of other users. Further research on \red\ 
\cite{Mieghem2011a}  showed a non trivial dependency between likes and dislikes
at the collective level, in line with the questions we address in this article.

\paragraph{Online polarization}
The proliferation of online participatory media, such as social network sites,
blogs and online fora, increases users engagement in discussions on political
and societal issues, which in its turn may - under certain conditions - split
individuals apart in their opinion space. Opinion polarization is
characterzied  by a division of the population into a small number of
fractions with high  internal consensus and sharp disagreement between them
\cite{Flache2011}.  Agent-based models \cite{Lorenz2007,Mas2013} and
experimental studies \cite{Van2009}  explain some aspects of opinion formation
and its role in consensus and polarization.

Based on data from digital traces, previous research investigated polarization
from the network perspective  in political blogs  \cite{Adamic2005}, in
follower and mention links in \tw\ \cite{Conover2011a},  and in   Swiss
politicians profiles \cite{Garcia2015a}, as well as in a non-political domains
like friendship networks \cite{Guerra2013}, and cultural expression
\cite{Garcia2013}. Additionally, exprerimental evidence shows that  group
processes like polarization function differently in  computer-mediated
communication than in a face-to-face interaction \cite{Taylor2002}, for
example as the  relative annonymity of online media dampens inhibiting effects
like the spiral of silence \cite{Noelle1993}.

\paragraph{Online emotions}

Emotional expression through online text has been analyzed in earlier research
on data from MySpace \cite{Thelwall2010b}, Yahoo answers \cite{Kucuktunc2012},
IRC channels \cite{Garas2012}, Wikipedia \cite{Iosub2014}, BBC, Digg, YouTube
and Twitter \cite{Thelwall2012}. Availability  of large-scale quantitative
datasets allows us to understand emotions  and their role in various domains.
Studies in the field of subjective well-being  leverage extensively on
quantifying emotions through text. For  instance, subjective well-being is
manifested in \fb\ status updates  \cite{Wang2014}, and shows a pattern of
assortativity in social networks  \cite{Quercia2012} in relation to feelings
of loneliness \cite{Burke2010}. This  is in a close relation to the
quantification of mood in \tw\, which has been  used to validate theories
of periodic mood oscillations \cite{Golder2011}.  

\tw\ mood
measured in terms of valence and arousal reveals  aspects of the relation
between mood states and online interaction and  participation
\cite{DeChoudhury2012}, and the psycholinguistic analysis of emotions reveal
the traces of mental health issues \cite{DeChoudhury2014}.
Furthermore, segregation patterns in geographical space 
\cite{Lin2014} and gender-based patterns \cite{Kivran2012, Thelwall2010b} can be 
partially attributed to differences in emotional expression. In online 
interaction, for example in real-time chat conversations \cite{Garas2012} and 
product reviews \cite{Garcia2011}, emotions are not a phenomenon characteristic 
to just an individual, but exhibit collective properties \cite{Schweitzer2010}.

Lastly, information-centric role of online emotions has been studied through
blogs \cite{Miller2011}, in \tw\ \cite{Pfitzner2012}, and in \texttt{Yahoo} 
answers \cite{Kucuktunc2012}. Emotions are the building blocks for a creation of
social network structures \cite{West2014, Tan2011} through empathy 
\cite{Kim2012} that lead to correlations between emotional expression and
popularity \cite{Kivran2011, Tchokni2014}. Negative emotional posts were
shown to be  drivers of communication among users and responsible for
extension of the  lifetime of online discussions in forums \cite{Chmiel2011b}.
Furthermore, the digital traces of emotions synchronize with political
outcomes \cite{Gonzalez-Bailon2010}, which goes inline with the findings that
political discussions are emotionally charged \cite{Hoang2013}, in particular
during  election periods \cite{Schweitzer2012}.

\section{Data and Methods}

\subsection{Data on collective evaluations}

\paragraph{Datasets}

The data used in this research is the result of our crawl of four  publicly
accessible online communities.

\yt\ (\url{http://www.youtube.com/}) is a video sharing website on which
registered users can upload and view videos, as well as post comments and rate
videos with likes and dislikes. Our crawl\footnotemark[3] was launched in
June 2011 to daily collect a combination of top videos in various categories
and to iteratively explore the channels of general users \cite{Abisheva2014},
including 6.3 Million videos by February 2015.

\footnotetext[3]{YouTube Data API Java wrapper 
(\url{https://developers.google.com/youtube/v3/})}

\red\ (\url{http://www.reddit.com}) is a message board in which registered
users  submit posts with links and text, and vote up and down for posts to
appear on a frontpage. Conversations between users appear in one of the many
thematic boards, called subreddits, covering diverse topics from politics to
science fiction and adult content. From 2012 to 2014 our daily \red\
crawl\footnotemark[4] collected 338,000 submissions from 1,972 subreddits. While 
the user interface of \red\ provides fuzzed amounts of votes, it is possible to 
construct the total amount of up and downvotes to a submission based on the 
JSON fields of reddit score and like ratio. This way, we count with the text and 
the final amount of up and downvotes for each submission in our dataset.

\footnotetext[4]{PRAW (\url{https://pypi.python.org/pypi/praw})}

\img\ (\url{http://www.imgur.org/}) is an image hosting and sharing  website
where registered users upload, rate, and discuss uploaded images.  Image
sharing traffic of \img\ has a large presence in \red\, such that every  6th
successful \red\ post has a link to an image on \img\ \cite{Olson2015}. Our
daily crawl\footnotemark[5] collected 200,000 images and their user activity 
statistics between December 2015 and January 2016. 

\footnotetext[5]{PyImgur Python API wrapper
(\url{https://github.com/Damgaard/PyImgur}). Seed images were selected from
\img's gallery sitemap (\url{http://imgur.com/gallery/sitemap.xml})}

Finally, \ud\ (\url{http://www.urbandictionary.com/}) is  an online
crowdsourced platform consisting of non-standard lexicon of slang words and
idioms. Registered users can submit new terms and provide definitions, and all
users of the website, registered and anonymous, can vote up and down for the
best definitions. Between April  and May 2013 our python-based crawl collected
220,000 definitions and their votes.

All platforms provide functionality for users to evaluate uploaded  content 
positively and negatively by clicking an upvote/\like\ or  downvote/\dislike\
button respectively. For simplicity, from now on we refer to evaluated videos,
submissions, images and definitions as \emph{items} and we denote as likes and
dislikes to positive and negative evaluations, including up and down votes
respectively.

\begin{table}[t]
 \centering
 \renewcommand{\tabcolsep}{1mm}
 \resizebox{\textwidth}{!}{
 \begin{tabular}{l D{.}{.}{10.1} D{.}{.}{11.1} D{.}{.}{11.1} D{.}{.}{13.1}
D{.}{.}{13.1}}
 \toprule
 \multirow{2}{*}{Dataset}& \multicolumn{3}{c}{Number of items, $N$} &
\multirow{2}{*}{Num. of likes}& \multirow{2}{*}{Num. of dislikes}\\
 \cline{2-4}
 &N_{\text{crawled}}&N_{\text{year} \geqslant 1}& N_{\text{L,D} \geqslant
1}&&\\
 \toprule
  \texttt{Urban Dict.} 
definitions&220,270&213,512&208,441&61,100,699&26,508,869\\
  \yt\ video descriptions&6,279,461&3,864,480&2,750,554&763,291,676&41,214,035\\
  \red\ submissions&338,845&174,444&142,662&5,078,242&947,519\\
  \img\ image titles&201,181&147,752&125,230&54,786,629&1,931,918\\
 \bottomrule
 \end{tabular}
 }
 \vspace{-7pt}
 \caption{\textbf{Number of items in each dataset}. $N_{\text{crawled}}$
counts the number of crawled items, and $N_{\text{year} \geqslant 1}$ the 
number of items in English \textit{and} that existed for more than a
year. $N_{\text{L,D} \geqslant 1}$ counts items that
received at least 1 like \textit{and} 1 dislike.}
 \label{tab:stats-num-posts}
 \vspace{-15pt}
\end{table}

\paragraph{Sentiment Analysis}

To quantify emotional expression, we applied sentiment analysis to headers or
titles of each item, leaving for a future research the analysis of longer
descriptions, transcripts, and comments. We applied sentiment analysis
techniques to video descriptions in \yt, image titles in \img, submission 
headers in \red\ and term definitions in \ud. Headers and titles are a good 
proxy of the  emotional tone of a discussion, in line with earlier research on 
forum-like conversations \cite{Gonzalez-Bailon2010}.

We measured emotional content of items by applying two complementary sentiment
analysis methods.  First, we apply a lexicon of affective norms of  valence
\textbf{V}, arousal \arousal\ and dominance \dominance\ of nearly 14,000
English words \cite{Warriner2013}. In line with previous findings
\cite{Warriner2013},  the scores of valence and dominance in our dataset are
highly correlated,  in comparison with the weaker correlation between valence
and arousal as explained more in detail in the Results section. This motivates
our focus to only valence and arousal as suggested by the theory of core
affect.

Second, we apply the SentiStrength classifier \cite{Thelwall2010a,
Thelwall2012}  a state-of-the-art lexicon-based method \cite{Kucuktunc2012,
Abbasi2014} that has been used in earlier research on the online data from
MySpace \cite{Thelwall2010b}, Yahoo! \cite{Kucuktunc2012}, IRC channels
\cite{Garas2012}, BBC, Digg, YouTube \cite{Thelwall2012}, Twitter
\cite{Thelwall2012, Pfitzner2012} and Wikipedia \cite{Iosub2014}. The core of
the SentiStrength method is to predict the sentiment of a text, based upon the
occurrences of the words from a lexical corpora, which contains the set of
terms with known sentiment of a text.  The classifier incorporates various
rules, which strengthen or  weaken sentiments of the lexicon words detected in
the short text.  Among the rules are syntactic rules, e.g. exclamation marks
and punctuation,  language modifiers and intensifiers, such as negation and
booster words, and  spelling correction rules. The final sentiment score is
composed of  a positive \posit\ and a negative \negit\ score for each text as
two discrete values  in the range of $[+1,+5]$ and $[-5,-1]$ respectively. In
our analysis, we normalize all emotions variables to  $[0..1]$ mapping \posit\
from $[+1,+5]$ to $[0,1]$ and reversing and rescaling \negit\  from $[-1,-5]$
to $[0,1]$.

To ensure a valid measurement of sentiment and collective evaluations, we
apply two filters to our datasets. First, since both sentiment analysis
techniques are designed only for English texts, we apply language
classification \cite{Nakatani2010} and filter out all non-English texts.
Second, we remove all items with less than a like and a dislike, and that
existed for less than a year in all platforms, to ensure that positive and
negative evaluations are stable. Detailed statistics on the number of posts in
each dataset are shown in Table \ref{tab:stats-num-posts}, showing that they
are still sufficient for large scale analyses. We will make these datasets
available for research purposes.

\subsection{Statistical analysis methods}

\paragraph{Distribution fits}

We apply a Maximum Likelihood criterion to fit the distributions of likes and
dislikes \cite{Alstott2013}, to confirm early findings of the fits of the 
popularity distribution to the log-normal distribution \cite{Mieghem2011b, 
Asur2011}. We use the \texttt{powerlaw} python package to fit four statistical 
distributions related to complex growth phenomena \cite{Mitzenmacher2004}: power 
law, log-normal, truncated power law and exponential distributions. We compare 
the likelihood of each distribution using the log-likelihood ratio 
$R=ln(L_1/L_2)$ between the two  candidate distributions and its significance 
value $p$. Positive ratios indicate
evidence for the first distribution, and negative ratios for the second one.
Instead of testing the hypothesis of the data following a certain
distribution, this comparative  test answers the question of which parametric
distribution provides the best fit available, following the principle of
Maximum Likelihood estimation \cite{Alstott2013}. To finally assess the
quality of the best fit, we measure the Kolmogorov-Smirnov distance between
the best fitting distribution and the emprical data.

\paragraph{Dual regime analysis}

We test the existence of a dual local versus global regime in collective 
evaluations by analyzing the non-linear properties of the relationship between 
the amounts of likes and dislikes for each item.

We use an extension of a traditional linear modelling, multivariate adaptive
regression splines (MARS)  \cite{Friedman1991, Friedman1993} implemented in
the \textbf{\textsf{R}} programming language package \textit{earth}. MARS
fits a continuous piecewise regression function with \emph{knots} that join
locally linear pieces.  In our analysis, we  are interested to test a dual
pattern  in the relationship between the number of  likes L and the number of
dislikes D, therefore we set the number of knots to one and fit a model of
the form 
\begin{equation*}
\text{D(L)} = I + \alpha_1 * max(0,\text{L}-L_c) +  \alpha_2 *
max(0,L_c-\text{L})
\end{equation*}
The values of likes above $L_c$ \underline{\textit{and}}  the values of
dislikes above D($L_c$) correspond to observations in the global regime, after
the bubble bursts, and the values in which any is below map to the local regime.

To evaluate the quality of the MARS model, we compare it to the Ordinary Least
Squares (OLS) regression  using the  Generalized Cross-Validation prediction
error (GCV) defined as   
\begin{equation*}
GCV = \frac{RSS}{N*(1-\frac{\text{ENP}}{N})^2}
\end{equation*}
where $N$ is the number of observations, $RSS$ is the residual sum of squares,
and $ENP$ is the effective number of parameters to avoid overfitting
\cite{Friedman1993}. We use the implementation provided by the package 
\textit{boot} in \rlang\, as well as the  coefficient of determination $R^2$ of 
both OLS and MARS fits.

\paragraph{Emotion and polarization analysis}

Having identified the two regimes and their thresholds in the relationship
between  the number of dislikes and the number of likes, we can mark items
either in the global or the local regime as a binary class. We test how
emotions influence the chances of items reaching the global regime through two
logistic regression models, one for each sentiment analysis technique.
Similarly, we combine the values of likes and dislikes through their geometric
mean to measure polarization, as manifested by simultaneous large amounts of
positive and negative evaluations. We regress this measure of polarization
through two linear models depending on the emotions expressed on the items.

Prior to modelling, we examine the normalized emotional  dimensions for
multicollinearity by computing the Spearman's rank correlation  coefficients,
to avoid singularities. We assess the quality of fits in comparison to null
models, by measuring the $\chi^2$ statistic of model likelihood ratio tests
implemented in the \emph{lmtest} \textbf{\textsf{R}} package.

\newpage

\section{Results}

\subsection{Stylized facts of evaluation distributions}

\begin{figure}[hb]
  \centering
  \includegraphics[width=0.95\textwidth]{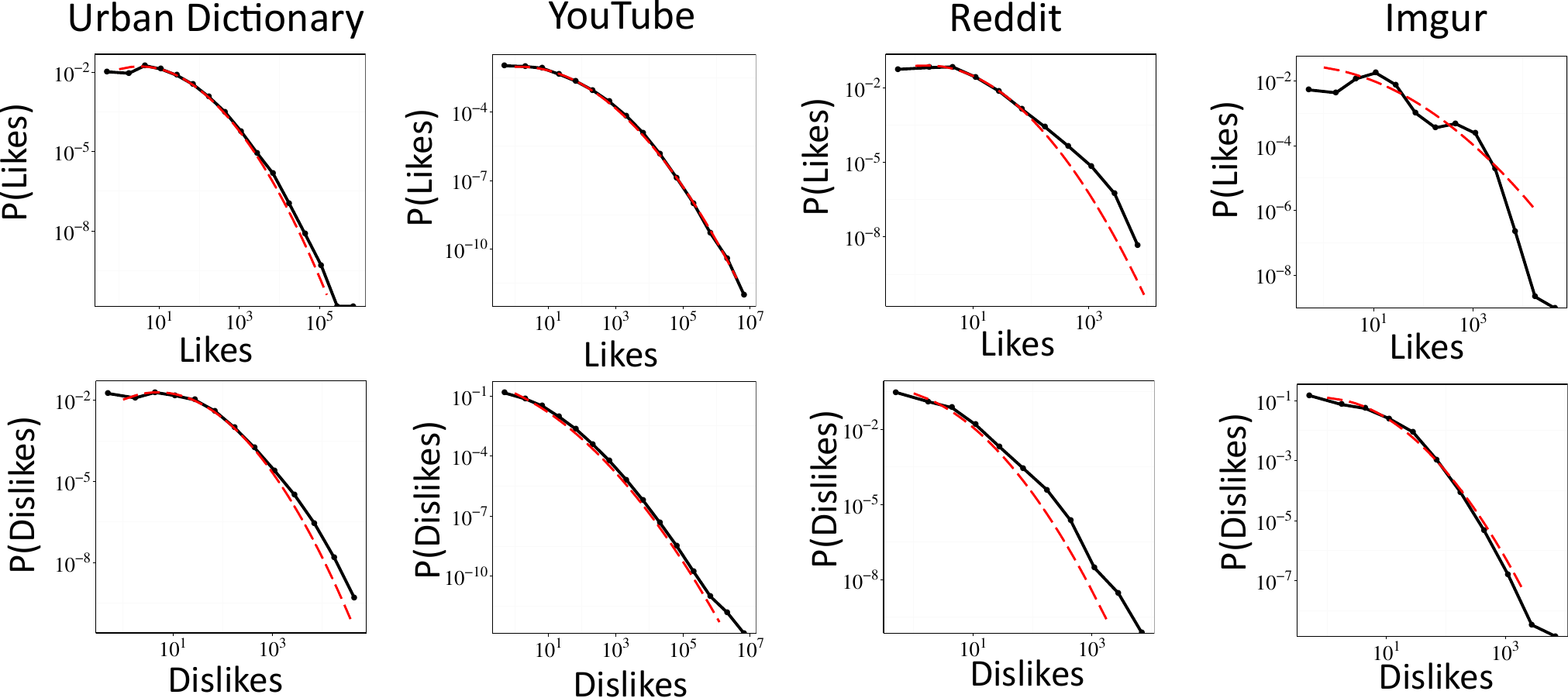}
  \vspace{-5pt}
  \caption{\textbf{Probability density function of collective evaluations}.
Probability density function of the number of likes (top) and the number
of dislikes (bottom) with exponential binning and fits to log-normal
distribution $ln\mathcal{N}(\mu,\sigma)$ (red dashed lines). For all datasets, 
the results of the log-likelihood pairwise comparisons of the four 
distributions 
(see text) identified the log-normal distribution as the best fit.}
  \label{fig:04-ccdfFitsLNLD}
\end{figure}

Figure \ref{fig:04-ccdfFitsLNLD} shows the probability density functions of
the distributions of the amount of likes and dislikes for items in each of the
four datasets.   To understand the process that generates these distributions,
we fit a set of parametric distributions that provide insights into how likes
and dislikes are given to items. Following the categorization of
\cite{Mitzenmacher2004}, generative mechanisms produce stylized size
distributions that can be traced back to the properties of growth processes.
If the appearance of likes and dislikes follows an uncorrelated process and
new evaluations are independent of previous ones, likes and dislikes should
follow \emph{exponential} distributions. On the other hand, the presence of
likes and dislikes can motivate further evaluations through social effects,
creating multiplicative growth (also known as preferential attachment in the
context of networks). In the presence of multiplicative growth, if items have
similar lifespans,   likes and dislikes follow \emph{log-normal} distributions. 
On the other hand if multiplicative growth is combined with heterogeneous 
lifespans, likes and dislikes follow a \emph{power law} distribution. This power 
law can be corrected by adding an exponential cutoff if finite size effects 
limit the growth of likes and dislikes, a case in which the distributions would 
be better fitted by a \emph{truncated power law}.

\begin{table}[tp]
 \resizebox{\textwidth}{!}{
  \begin{tabular}{c c c c c c c c c}
  \hline
  &\multicolumn{2}{c}{\ud}
  &\multicolumn{2}{c}{\yt}
  &\multicolumn{2}{c}{\red}
  &\multicolumn{2}{c}{\img}\\
  
&P(Likes)&P(Dislikes)&P(Likes)&P(Dislikes)&P(Likes)&P(Dislikes)&P(Likes)& 
P(Dislikes)\\
  \hline
  $\mu$    & 4.092 & 3.657 & 5.492 & 1.405 & 2.197 & 0.492 & 4.668 & 1.821\\
  $\sigma$ & 1.705 & 1.435 & 2.28  & 2.528 & 1.332 & 1.35  & 2.46  & 1.447\\
  $D$      & 0.008 & 0.008 & 0.009 & 0.002 & 0.029 & 0.01  & 0.098 & 0.031\\
  \hline

$ln \left( \frac{L_{\text{LN}}}{L_{\text{PL}}}\right)$ 
	& $123115.6^{***}$ & $129462.1^{***}$ & $174835.6^{***}$ & 
$42309.5^{***}$ 
	& $45355.3^{***}$  & $13898.9^{***}$  & $65314.8^{***}$  & 
$34103.6^{***}$\\
  
$ln \left( \frac{L_{\text{LN}}}{L_{\text{TPL}}}\right)$
	& $55538.6^{***}$  & $61135.1^{***}$  & $252646.1^{***}$ & 
$42592.3^{***}$
	& $17182.5^{***}$  & $10084^{***}$    & $20108.5^{***}$  & 
$3662.4^{***}$\\
  
$ln \left( \frac{L_{\text{LN}}}{L_{\text{EXP}}}\right)$
	& $260098.8^{***}$ & $126865^{***}$ & $953075^{***}$ & $1270237.6^{***}$
	& $149405^{***}$   & $70261.5^{***}$& $75817.7^{***}$& $26681.4^{***}$\\
  \hline
  \end{tabular}}
  \vspace{-5pt}
  \caption{\textbf{Log-normal fit parameters of collective evaluations and 
comparison to other distributions}. Estimated parameters of the fitted
log-normal distribution $ln\mathcal{N}(\mu,\sigma)$ and Kolmogorov-Smirnov
distances $D$. The bottom row shows the log-likelihood ratios of pairwise
comparison between the log-normal distribution fit (numerator) and the other
three distributions: power law, truncated power law and exponential. All three
ratios are positive, large and significant ($p < 0.05$) which confirms that 
among the four candidate distributions the log-normal distribution is the best 
fit.}
  \label{tab:03-ln-ks}
  \vspace{-10pt}
\end{table}

For all datasets, the results of pairwise comparisons of the four proposed
distributions identified the \textit{log-normal} distribution as the best fit,
with significant and positive log-likelihood ratios as shown in  Table
\ref{tab:03-ln-ks} along with the best fitting parameter estimates. The dashed
lines in Figure \ref{fig:04-ccdfFitsLNLD} show the fitted distributions,
revealing the quality of the fit. The cases of  \yt\ and \ud\ provide very good 
fits with extremely low Kolmogorov-Smirnov $D$ statistics.  The fits are not so 
good at the tails of \red\ and \img, but the the Kolmogorov-Smirnov $D$ 
statistic provide good values below 0.05 and the \textit{log-normal} 
distribution clearly outperforms all others. The worst fit is for the  number of 
likes in \img, for which Figure  \ref{fig:04-ccdfFitsLNLD} suggests a bimodal
pattern. Identifying the possible mechanisms that can produce such bimodality  
goes beyond the scope of this research. We can conclude that the amounts of 
likes and dislikes display a general heavy tailed behavior of 
\textit{log-normal} distributions, lending evidence for the production of 
evaluations following socially coupled growth processes with homogeneous life 
spans.

\subsection{The dual pattern of collective evaluations}

\begin{figure}[t]
  \centering
  \includegraphics[width=0.95\textwidth]{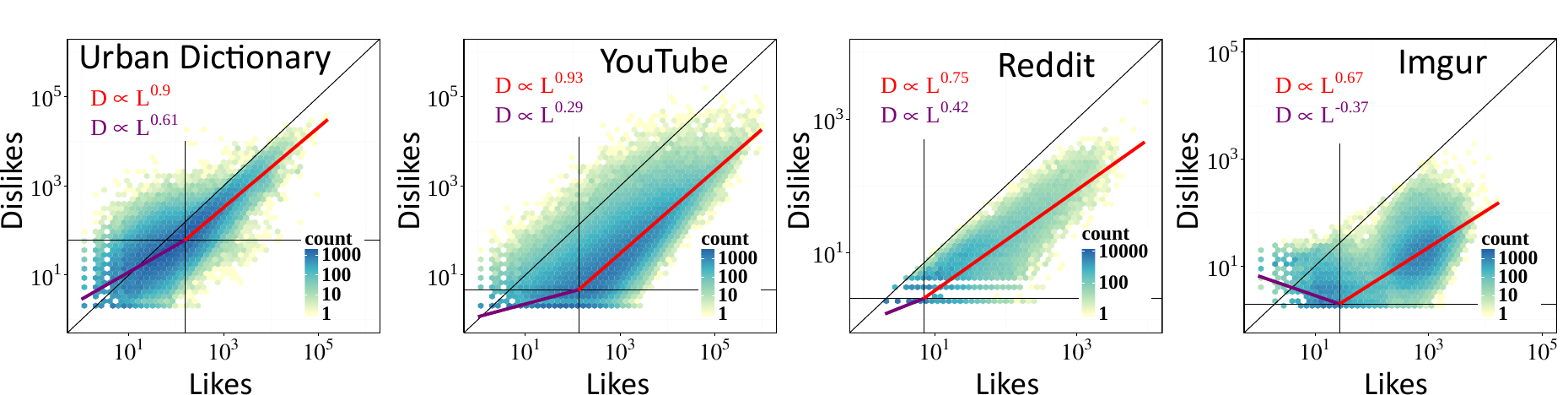}
  \caption{\textbf{Relationship between the number of dislikes and likes}.  
Two-dimensional joint distributions with 50 bins, bin colors indicate the  
count of observations within the bin. Purple and red lines show the local and 
global regimes of the non-linear relationship between the number of dislikes 
and 
the number of likes.  Threshold estimates are located at $L_c$, estimated  as 
$L_{c} = 155$ in \ud; $L_{c} = 131$ in \yt; $L_{c} = 7$ in \red; and $L_{c} = 
27$ in \img.}
  \vspace{-10pt}
  \label{fig:03-2DhistLDplot}
\end{figure}

We explore the existence of a dual relationship between likes and dislikes
through non-linear MARS fits, testing if the relationship can be divided in a 
local and a global regime. We  restrict the number of model terms to have a
single knot, measuring if a dual model outperforms a linear pattern. Figure
\ref{fig:03-2DhistLDplot} shows the results of MARS fits between the logarithms 
of likes and dislikes. Vertical and horizontal lines mark the likes cutoff value 
$L_c$ and its corresponding value of dislikes in the fit D($L_c$). These cutoff 
values divide the system in a local versus a global regime, with the fitted 
functions of the form $D \propto L^{\lambda}$ and $D \propto L^{\gamma}$ 
respectively.

In all datasets, the exponent of the global regime is larger  than exponent of
the local one, for example in \yt\ $\gamma = 0.93 >  \lambda = 0.29$. While both 
exponents are below $1$ and indicate sublinear scaling, the much higher value of 
the second one shows that, beyond a threshold value of likes, the dislikes given 
to items grow faster than below the threshold as a sign of the burst of a 
filter bubble. The presence of scaling in \red\ votes was previously  reported 
in a smaller data subsample \cite{Mieghem2011a}, concluding the existence of 
superlinear scaling of dislikes with likes. Our analysis shows that the 
relationship between likes and dislikes in \red\ is better approximated 
by a dual regime model, in line with the results of the other three datasets.

We evaluate the goodness of the dual model against a single regime model in 
Table \ref{tab:03-r2-lm-mars}. The dual model outperforms in $R^2$ and GCV to 
the single regime model, lending strong evidence to the existence of two
regimes. We further tested if additional knots could improve the fits, and
found that a dual regime is the optimal model for \ud, \yt, and \red, and only a 
4 knot model could improve the \img\ fit by a marginal GCV of less than 0.01. 

\begin{table}[h]
\centering
\resizebox{0.7\textwidth}{!}{
\begin{tabular}{l D{.}{.}{4.3} D{.}{.}{5.3} D{.}{.}{5.3} D{.}{.}{5.3} }
\hline
model & \multicolumn{1}{c}{\texttt{Urban Dict.}} & \multicolumn{1}{c}{\yt} 
& \multicolumn{1}{c}{\red} & \multicolumn{1}{c}{\img} \\
\hline
$R^2$ (\texttt{lm}) &0.646&0.634&0.727&0.505\\
$R^2$ (\texttt{MARS}) &0.654&0.683&0.741&0.597\\
\hline
$GCV$ (\texttt{lm}) &0.785&1.283&0.301&0.804\\
$GCV$ (\texttt{MARS)}&0.767&1.111&0.286&0.654\\
\hline
\end{tabular}}
  \caption{\textbf{The goodness of the dual and the linear model}. Comparison of
the linear and the MARS models of the relationship between the number of 
dislikes and the number of likes. Top row shows the coefficient of determination 
$R^2$ (higher is better). Bottom row shows the generalized 10-fold 
cross-validation prediction error (GCV) (lower is better). The dual model 
outperforms in $R^2$ and in GCV compared to the linear model.}
\label{tab:03-r2-lm-mars}
\vspace{-20pt}
\end{table}

\subsection{Emotions in the global regime}

\begin{figure}[t]
 \resizebox{\textwidth}{!}{
   \centering
   \begin{subfigure}[b]{0.245\textwidth}
     \includegraphics[width=\textwidth]{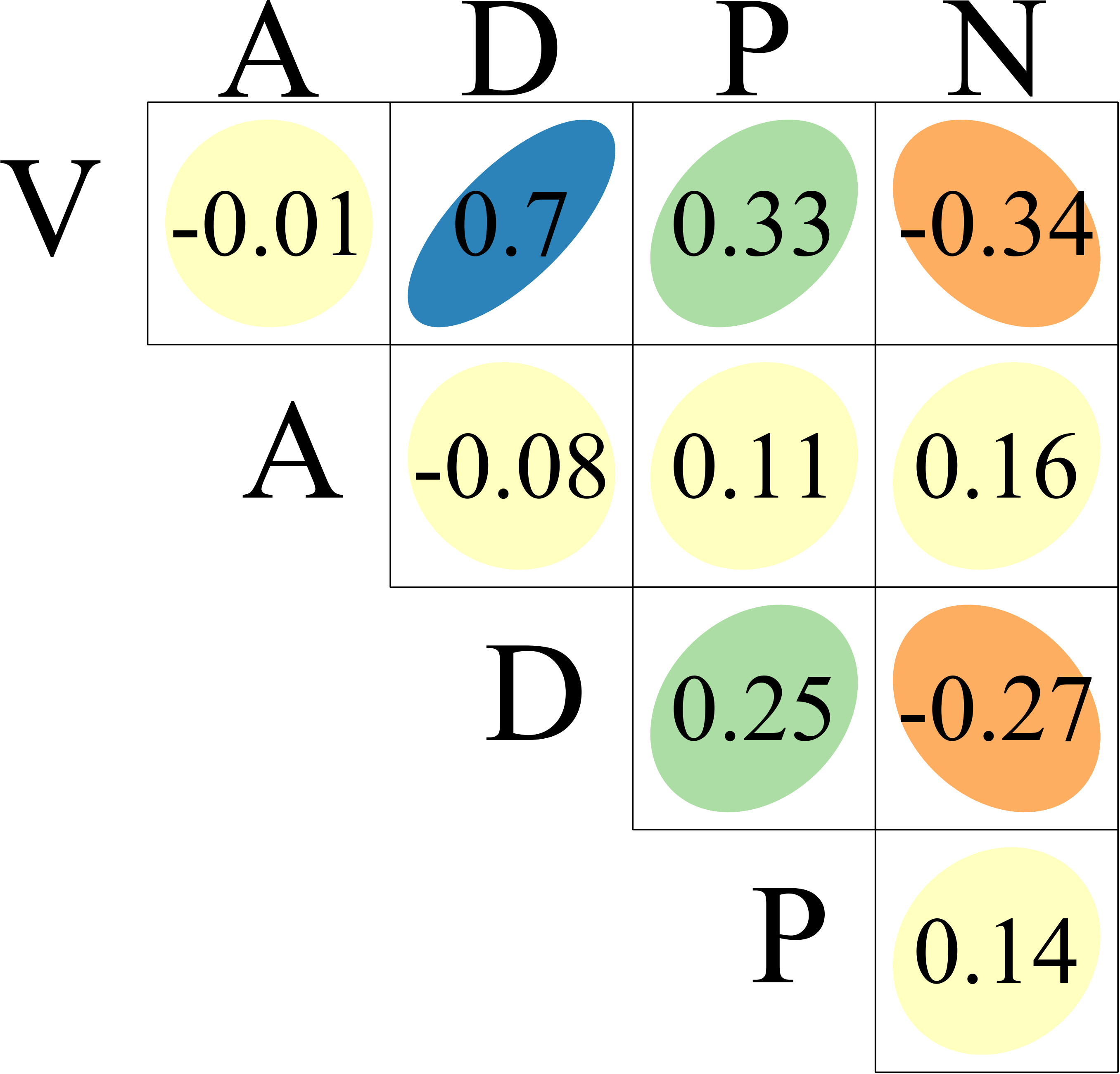}
   \end{subfigure}\hspace{-1pt}
   \begin{subfigure}[b]{0.245\textwidth}
     \includegraphics[width=\textwidth]{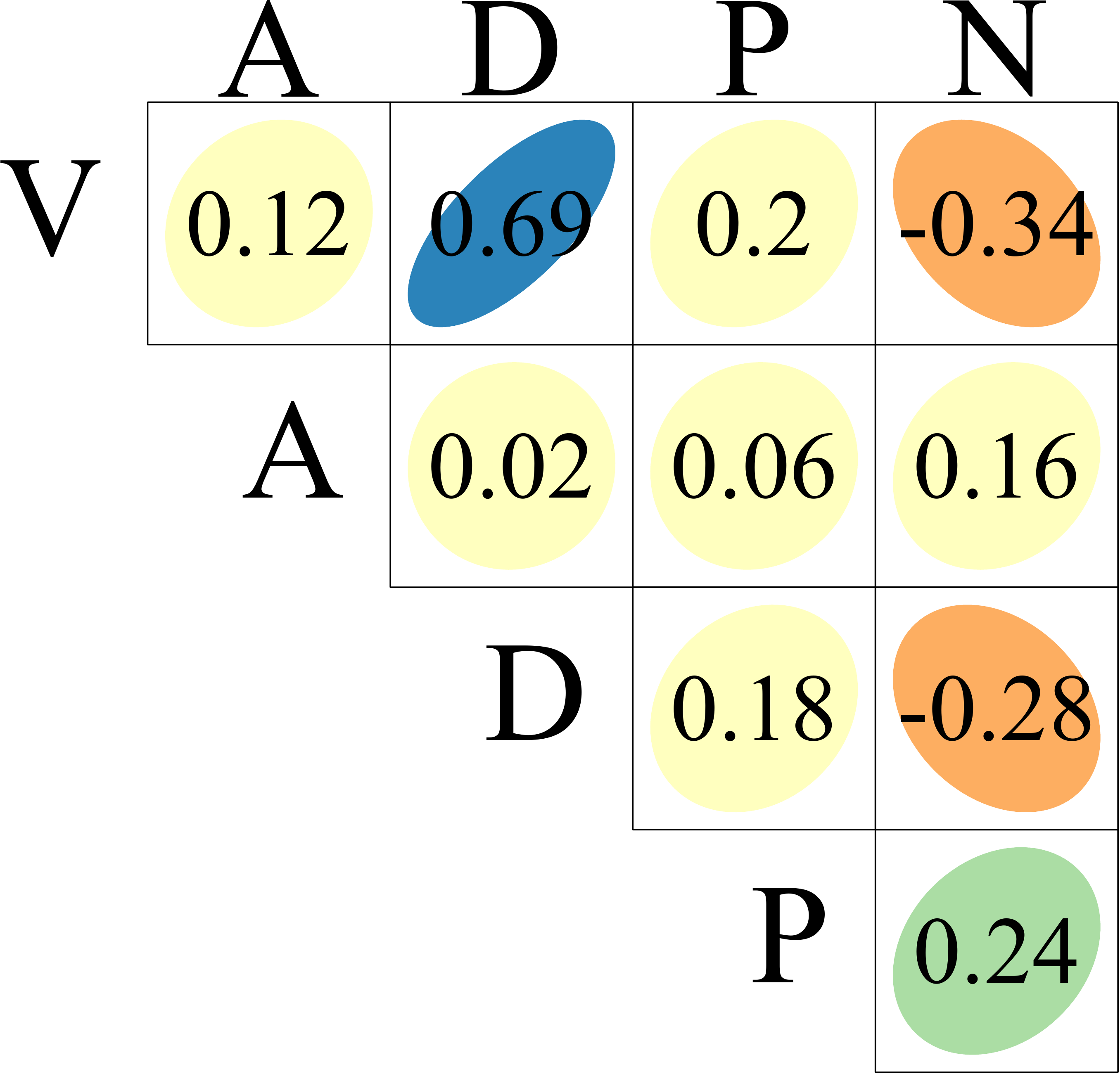}
   \end{subfigure}\hspace{-1pt}
   \begin{subfigure}[b]{0.245\textwidth}
     \includegraphics[width=\textwidth]{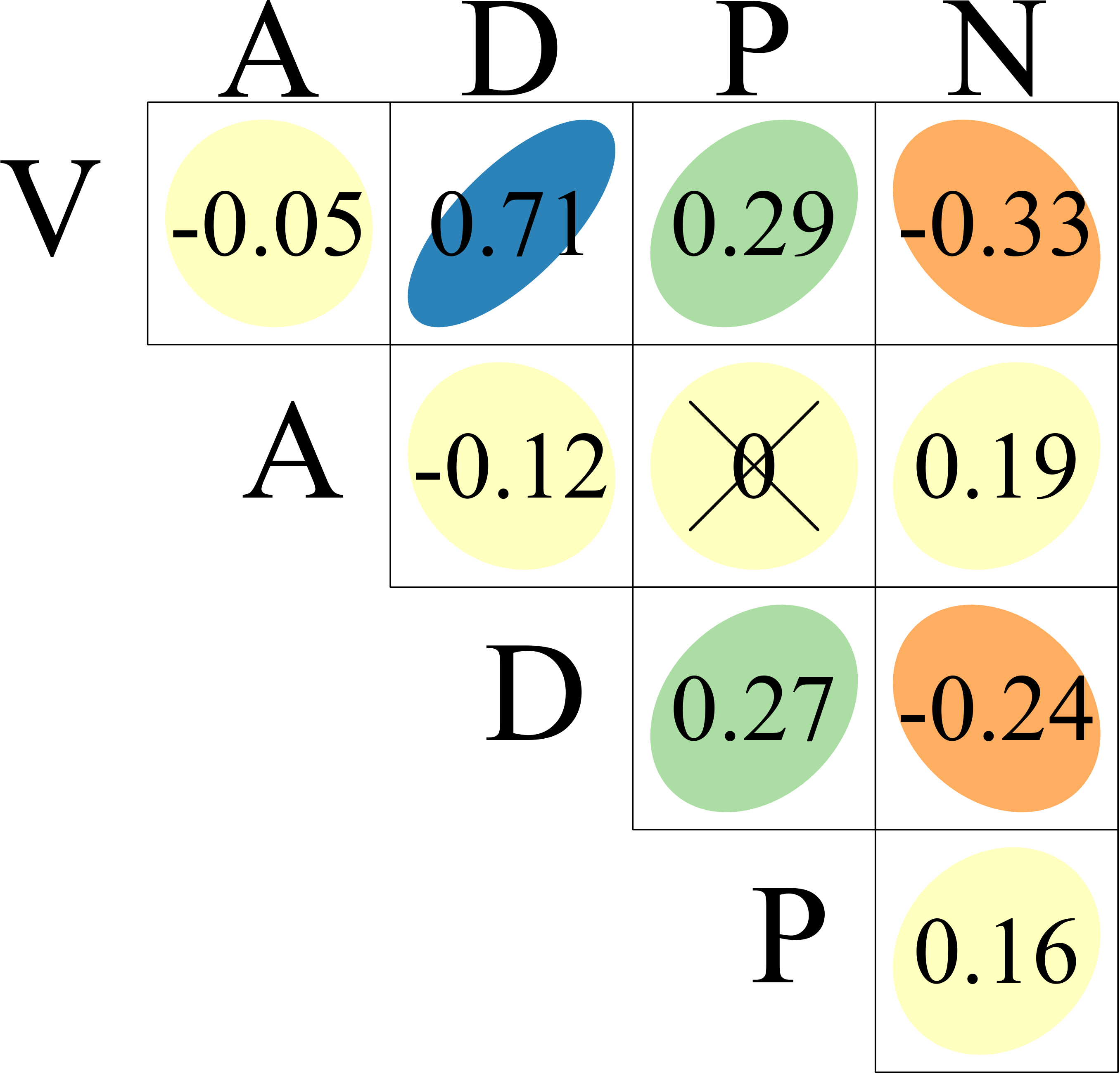}
   \end{subfigure}\hspace{-1pt} 
   \begin{subfigure}[b]{0.28\textwidth}
     \includegraphics[width=\textwidth]{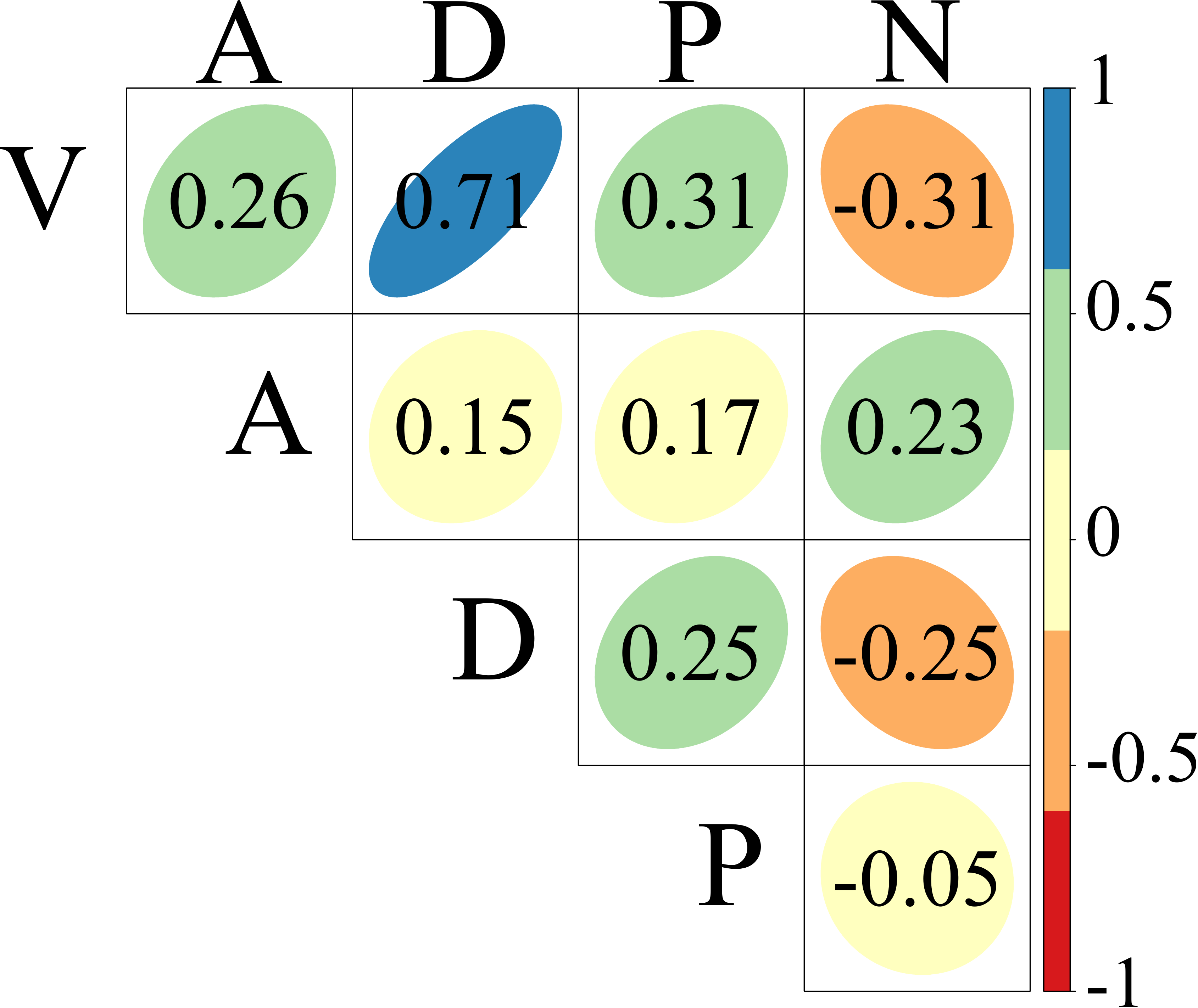}
   \end{subfigure}
  }
   \caption{\textbf{Correlations of emotions}. Spearman's correlation matrix of
emotional dimensions, in an order from left to right: A) Urban  Dictionary, B) 
YouTube, C) Reddit, D) Imgur. Significance level $p < 0.05$. Insignificant 
correlations are crossed out. Predictors are normalized to  $[0..1]$. Dominance 
is highly correlated with valence, and therefore the dominance variable is 
discarded from the further analysis.}
   \label{fig:07-emoscorrplot}
   \vspace{-15pt}
 \end{figure}

Figure \ref{fig:07-emoscorrplot} illustrates the rank correlations   between
emotional dimensions.  In all datasets  valence and dominance are
\textit{highly correlated} with $\rho \geqslant  0.7^{***}$, and therefore we
discard the dominance variable from regression  analysis as it is difficult to
distinguish from valence. Valence and positivity \posit\ have a minor positive
significant correlation $\rho \in [0.2,0.3]$, and valence and negativity \negit\ 
have a slightly  negative correlation $\rho \approx -0.3$, illustrating the 
relation of emotion variables accross both valence/arousal and positive/negative models.

We fit two regression models in which the probability of the event of an item
reaching the global regime $G$  depends on the emotions expressed in the
evaluated item.  The first model uses \valence\ and  \arousal\ as explanatory
variables, and focuses on the role of emotions as quantified through their
pleasant/unpleasant and active/calm dimensions. The second model takes \posit\
and  \negit\ as predictors, and measures significance of positive and negative
sentiments in bringing an item to global regime.  Table \ref{tab:07-lr-pn-pol}
reports the results of logistic regression of the form $logit(G) \sim V + A$
and  $logit(G) \sim P + N$ respectively.  The role of arousal is heterogeneous, 
having a significant positive effect in \ud\ and \img, but a weak negative 
effect in \yt\ and a non-significant one in \red. The effect of valence is also 
mixed, in \ud\ and \yt\ the chances of reaching the global regime grow with 
valence, while  in \red\ and \img\ is the opposite case. The second model sheds 
more light to this: the pattern is the same for positive sentiment, but 
negative sentiment increases the chance of reaching the global regime in all 
datasets but \red, where the effect is not significant.

\begin{table}[h]
\centering
\resizebox{0.7\textwidth}{!}{
\begin{tabular}{l D{.}{.}{4.3} D{.}{.}{7.3} D{.}{.}{5.3} D{.}{.}{5.3} }
\hline
  & \multicolumn{1}{c}{\texttt{Urban Dict.}} & \multicolumn{1}{c}{\yt} & 
\multicolumn{1}{c}{\red} & \multicolumn{1}{c}{\img} \\
\hline
Intercept            & -2.071^{***} & -0.305^{***} & -0.111^{***} & 0.228^{***} 
 \\
$V$                  & 0.976^{***}  & 0.618^{***}  & -0.262^{***} & 
-0.209^{***} \\
$A$                  & 0.584^{***}  & -0.049^{**}  & -0.006(n)       & 
0.300^{***}  \\
\hline
$\chi^2$  & 547.3^{***}      & 2791.4^{***}     & 44.1^{***}       & 35.7^{***} 
      \\
\hline
\hline
 & \multicolumn{1}{c}{\texttt{Urban Dict.}} & \multicolumn{1}{c}{\yt} & 
\multicolumn{1}{c}{\red} & \multicolumn{1}{c}{\img} \\
\hline
Intercept            & -1.369^{***} & -0.115^{***} & -0.259^{***} & 0.261^{***} 
 \\
$P$                & 1.019^{***}  & 0.581^{***}  & -0.166^{***} & -0.296^{***} 
\\
$N$                & 0.170^{***}  & 0.218^{***}  & -0.006(n)       & 
0.191^{***}  \\ 
\hline
$\chi^2$  & 2552.6^{***}     & 17150.9^{***}    & 41.9^{***}       & 
120.3^{***}      \\
\hline
\multicolumn{5}{l}{\scriptsize{$^{***}p<0.001$, $^{**}p<0.01$, $^*p<0.05$, 
$(n)$ not significant.}}
\end{tabular}}
\vspace{-5pt}
\caption{\textbf{The role of emotions in the global regime}. Logistic
regression models, $logit(G) \sim V + A$ and $logit(G) \sim P + N$, results for
probability of an item to be in a global regime. The effect of arousal and
valence is heterogeneous, and depends on the dataset.}
\vspace{-15pt}
\label{tab:07-lr-pn-pol}
\end{table}

\newpage

\subsection{Analysis of emotions in polarization}

Since the distributions of likes and dislikes are approximately log-normal, we
can  treat the logarithms of likes  $ln(L)$ and dislikes $ln(D)$ as centrally
distributed around their means  $\mean{ln(D)}$ and $\mean{ln(L)}$.
We standardize the logarithmic counts  $ln(D)$ and $ln(L)$ as:
\begin{equation*}
Z_L = \frac{ln(L) - \mean{ln(L)}}{sd(ln(L))} \quad Z_D = \frac{ln(D) -
\mean{ln(D)}}{sd(ln(D))}
\end{equation*}
 where $sd(ln(L))$ and $sd(ln(D))$ are the standard deviations. Then, we compute 
a measure of polarization as the geometric mean of both values: 
$Pol=\sqrt{Z_L*Z_D}$. This measure captures the principle that polarization is 
high under simultaneous large amounts of positive and negative evaluations, and 
that polarization is low when only one of the values is dominant.

To understand which kind of emotional content creates polarization, we fit two
regression models as in the previous section, one of polarization as a function 
of valence and arousal in the evaluated item, and another as a function of 
positive and negative sentiment scores. The results of the fits are shown in 
Table \ref{table:LikesModels}. In line with the theory that links arousal to 
more extreme opinions,  we find a general pattern in three datasets  where 
arousal leads to higher levels of polarization. While there is no significant 
effect in \red, all the other datasets show that items that contain words that
transmit higher arousal also create a stronger polarized response.

\begin{table}[bh]
\centering
\resizebox{0.7\textwidth}{!}{
\begin{tabular}{l D{.}{.}{4.3} D{.}{.}{7.3} D{.}{.}{5.3} D{.}{.}{5.3} }
\hline

  & \multicolumn{1}{c}{\texttt{Urban Dict.}} & \multicolumn{1}{c}{\yt}   &
 \multicolumn{1}{c}{\red} & \multicolumn{1}{c}{\img}  \\

\hline
Int.           & 2.0508^{***} & 1.4543^{***}  & 1.3511^{***} & 1.7623^{***}  \\
$V$              & 0.3132^{***} & 0.2980^{***} & -0.1954^{***}  & -0.1908^{***} 
\\
$A$                     & 0.2662^{***} & 0.1005^{***} & -0.0327 (n) & 
0.2399^{***} \\
\hline
$\chi^2$    & 480.64^{***}   & 3420.4^{***} & 97.315^{***}  & 88.669^{***}      
  \\
\hline
\hline
  &  \multicolumn{1}{c}{\texttt{Urban Dict.}} & \multicolumn{1}{c}{\yt}  & 
  \multicolumn{1}{c}{\red}  & \multicolumn{1}{c}{\img} \\
\hline
Int.           & 2.2744^{***} & 1.5896^{***} & 1.2220^{***}  & 1.7625^{***}   \\
$P$               & 0.4889^{***} & 0.3059^{***} & -0.1107^{***}  & 
-0.1484^{***}  \\
$N$                  & 0.1194^{***} & 0.1698^{***} & 0.0077 (n)  & 0.1672^{***} 
\\ 
\hline
$\chi^2$     & 3271.2^{***}  & 19830.0^{***} & 63.073^{***} & 170.81^{***}      
 
 \\
\hline
\multicolumn{5}{l}{\scriptsize{$^{***}p<0.001$, $^{**}p<0.01$, $^*p<0.05$}, 
$(n)$ not significant}
\end{tabular}}
\caption{\textbf{The role of emotions in the polarization}. Linear regression
models, Pol $\sim V + A$ and Pol $\sim P + N$, results for polarization of the 
evaluation of an item as a function of emotions expressed on its text. Arousal 
and negativity drive polarization in all datasets except \red. The effect of 
valence and positivity is dataset-dependent.}
\label{table:LikesModels}
\end{table}

This also manifests in the model using positive and negative scores, where
negative content predicts higher polarization in the same three cases as for
arousal. The results of these two metrics are consistent with the hypothesis
that the expression of activating and negative feelings, such as anger or
outrage, tend to create more polarized responses, in line with the theoretical
argument that poses emotions as mechanisms to speed up evaluation processes at
the expense of more extreme reactions.

Valence in evaluated items creates different responses. Two communities,
\img\ and \red, show a negative relation of polarization with valence and
positive sentiment. The other two, \ud\ and \yt, show the opposite, where
polarization increases with valence. This  suggests  a context dependent
interpretation of positive expression, which does not necessarily motivate
positive empathy  but can also fuel polarized responses. The positive and
negative scores model works better than the valence and arousal model in all
cases but \red, where the valence and arousal model was more explanatory for
polarization, as evidenced by $\chi^2$ tests comparing both models.

\section{Discussion}

Our study of emotions focuses on understanding the role of emotions expressed
in the text of items with relation to the chances that the items reach the
global regime and produce polarized evaluations. While we used two established
and validated sentiment analysis methods based on metrics from psychology,
future advanced techniques can reveal new patterns and potentially falsify the
conclusions of our analysis with current techniques. Furthermore, deeper
analyses on individual data can correlate the expression of individual
emotions in the comments of a user and the evaluations given by the user,
bridging closer this way the measurement of emotional states and evaluations
and providing a better understanding of interpersonal emotions.

Following an observational approach to collective evaluations has the
advantage of having high ecological validity, but lacks the level of control
that can be induced in experimental scenarios. We can deduce insights on the
factual properties of collective evaluations, such as the dual regime between
likes and dislikes, but testing the conditions that produce them requires a
controlled set up.  Our motivation and explanation for the dual regime stems
from the phenomenon of filter bubbles \cite{Pariser2011}, but to fully
understand how these filters affect our behavior we need to experiment on how
individual evaluations respond to filtering mechanisms. While these
experiments can be carried out it in typical psychological settings and surveys,
large platforms like \fb\ can also experiment with the behavior
of their users in this respect (under the appropriate ethical considerations).
A complete understanding of online evaluations can only be achieved when our
results are complemented by experimental approaches.

The use of observational data has the advantage of taking a \emph{natural
exposure} approach: we analyze the evaluations of what people actually see,
rather than the \emph{forced exposure} to content  in experiments
\cite{McPhee1963}. In contrast, using digital traces of evaluations contains a
selection bias by which some users might be responsible for much larger
amounts of likes and dislikes than other users. While this selection bias is
natural at the collective level, inferring conclusions about the behavior of
individuals needs to consider corrections and use richer datasets
\cite{Cuddeback2004},  or apply agent-based modelling approaches to connect
the micro and macro levels \cite{Schweitzer2010}.

We explain the dual pattern between likes and dislikes as the result of filter
bubbles, but other possible explanations might also be plausible. Some unkown
deleting mechanism might downsample videos with a lot of dislikes in the local
regime, or some external factor like audience size might explain the  values
of the thresholds.  The results of our statistical analyses of distributions
of likes and dislikes fit to hypothetical mechanisms of multiplicative growth,
in line with previous findings on popularity metrics rather than evaluations
\cite{Mieghem2011b,Asur2011}. Our in depth statistics also provide a clear
view on the limits of our results, for example in the worse fits of log-normal
distributions in \img. Future research can conjecture on the possible
alternative explanations of our findings, in particular with respect to which
filtering mechanisms are in place. Our results do not allow us to distinguish
social filtering, based on friends and follower links, from recommender
systems, which are based on previous evaluations of a user. Further research
with information on individual behavior can shed light on these different
processes, for example measuring evaluation tendencies to content produced by
friends versus strangers, or across assortative and disassortative links with
respect to opinions.

Our analysis of the relation between likes and dislikes is based on the
amounts given to items after a long time has passed. This way, we evaluate
items after they do not attract lots of attention and their counts are stable.
In a figurative way, we study the \emph{fossils} of broken filter bubbles, but
we do not study them in a live setting. To fully understand the dynamics of
collective evaluations, we need data with temporal resolution on the counts of
likes and dislikes. In general, such data is not publicly available on the
sites, which requires a much more powerful crawling approach to monitor items
on a frequent basis, or access to proprietary data.


\section{Conclusions}


Our analysis of collective evaluations across various online media shows
statistical regularities in the distributions of evaluations and their
relationships. Our contribution is threefold:  First we report that the
distributions of the amounts of likes and dislikes per item are  well fitted
by log-normal distributions, a result that gives insights into the properties
of the process that creates evaluations. Second, we test the existence of a
dual pattern in the relation between likes and dislikes, finding robust
evidence of the existence of a local and a global regime that is consistent
with our hypotheses about the burst of filter bubbles.  Third, we found
evidence for the role of emotions in the creation of polarization and the
access to the global regime, lending support for psychology theories about the
role of affect, in particular arousal, in the polarization of opinions.


Our results have implications for the design of online platforms and filtering
mechanisms. Recommender systems and filtering mechanisms allow users to
discover content of relevance and quality, but can have unintended
consequences in the large scale. Our results suggest that the increasing
polarization levels of discussions might be created by these filtering
mechanisms, and that users are at risk of receiving a negative backlash to
their content when it goes beyond their local social context. Such abrupt
behavior with respect to negative evaluations can have important consequences
to user motivation and engagement, which might only be visible on the long
run.

Our findings shed light on fundamental polarization processes, in particular
with respect to the role of emotions. Increasing levels of polarization pose a
risk of social conflict and hinder collaboration and common goods, but a
healthy society needs certain level of disagreement to be able to deliberate,
discuss, and take decisions about important topics. Calibrating the design of
web and social media offers this way the chance to find a balance between
stagnation and polarization, leading to productive interaction in our current
online society.

\section{Acknowledgments:} 
This research was funded by the Swiss National Science Foundation 
(CR21I1\_146499/1).

\small
\raggedright
\sloppy
\bibliographystyle{acm}
\bibliography{websci2016}

\end{document}